\documentclass{article}

\PassOptionsToPackage{numbers, compress}{natbib}

\usepackage[final]{neurips_2022}

\usepackage[utf8]{inputenc} 
\usepackage[T1]{fontenc}    
\usepackage{hyperref}       
\usepackage{url}            
\usepackage{booktabs}       
\usepackage{amsfonts}       
\usepackage{nicefrac}       
\usepackage{microtype}      
\usepackage{xcolor}         
\usepackage{multirow}
\usepackage{graphicx}

\title{COVIDx CT-3: A Large-scale, Multinational, Open-Source Benchmark Dataset for Computer-aided COVID-19 Screening from Chest CT Images}

\author{
    Hayden Gunraj$^{1,2}$, Tia Tuinstra$^{1,2}$, Alexander Wong$^{1,2,3,4}$\\
    $^{1}$Vision and Image Processing Lab, University of Waterloo\\
    $^{2}$Department of Systems Design Engineering, University of Waterloo\\
    $^{3}$Waterloo AI Institute, University of Waterloo\\
    $^{4}$DarwinAI Corp.\\
    \texttt{\{hayden.gunraj, ttuinstra, a28wong\}@uwaterloo.ca}
}

\begin{document}

\maketitle

\begin{abstract}
    Computed tomography (CT) has been widely explored as a COVID-19 screening and assessment tool to complement RT-PCR testing. To assist radiologists with CT-based COVID-19 screening, a number of computer-aided systems have been proposed. However, many proposed systems are built using CT data which is limited in both quantity and diversity. Motivated to support efforts in the development of machine learning-driven screening systems, we introduce COVIDx~CT-3, a large-scale multinational benchmark dataset for detection of COVID-19 cases from chest CT images. COVIDx~CT-3 includes 431,205 CT slices from 6,068 patients across at least 17 countries, which to the best of our knowledge represents the largest, most diverse dataset of COVID-19 CT images in open-access form. Additionally, we examine the data diversity and potential biases of the COVIDx~CT-3 dataset, finding that significant geographic and class imbalances remain despite efforts to curate data from a wide variety of sources.
\end{abstract}

\section{Introduction}
    CT has shown great potential for detection of COVID-19, and in particular machine learning-based systems for detection of COVID-19 from CT images have been widely explored. Machine learning techniques allow for the visual characteristics of COVID-19 to be learned directly from CT images, which may aid clinicians in differentiating COVID-19 pneumonia from pneumonia of other etiology. However, even the largest studies in research literature have been limited in terms of quantity and/or diversity of patients, with many limited to single-nation cohorts~\cite{Mei2020,Gunraj2020,cncb,COVIDCTset,HUST,Xu2020,Ardakani2020,Javaheri2021,Wu2020,STOIC}. Multinational patient cohorts have been leveraged in several studies, but have typically been limited to few patients or few countries\cite{Hasan2020,Harmon2020,Jin2020_2}. In this study, we introduce COVIDx~CT-3, a large-scale multinational benchmark dataset for detection of COVID-19 cases from chest CT images comprising 431,205 CT slices from 6,068 patients across at least 17 countries. COVIDx~CT-3 builds upon COVIDx~CT-2~\cite{Gunraj2022}, and includes more than twice as many images. Additionally, we examine the data diversity and potential biases of the COVIDx~CT-3 dataset, finding that significant geographic and class imbalances remain despite efforts to curate data from a wide variety of CT data sources.

\section{Methods}

    \begin{figure}[!t]
        \centering
        \includegraphics[width=0.75\textwidth]{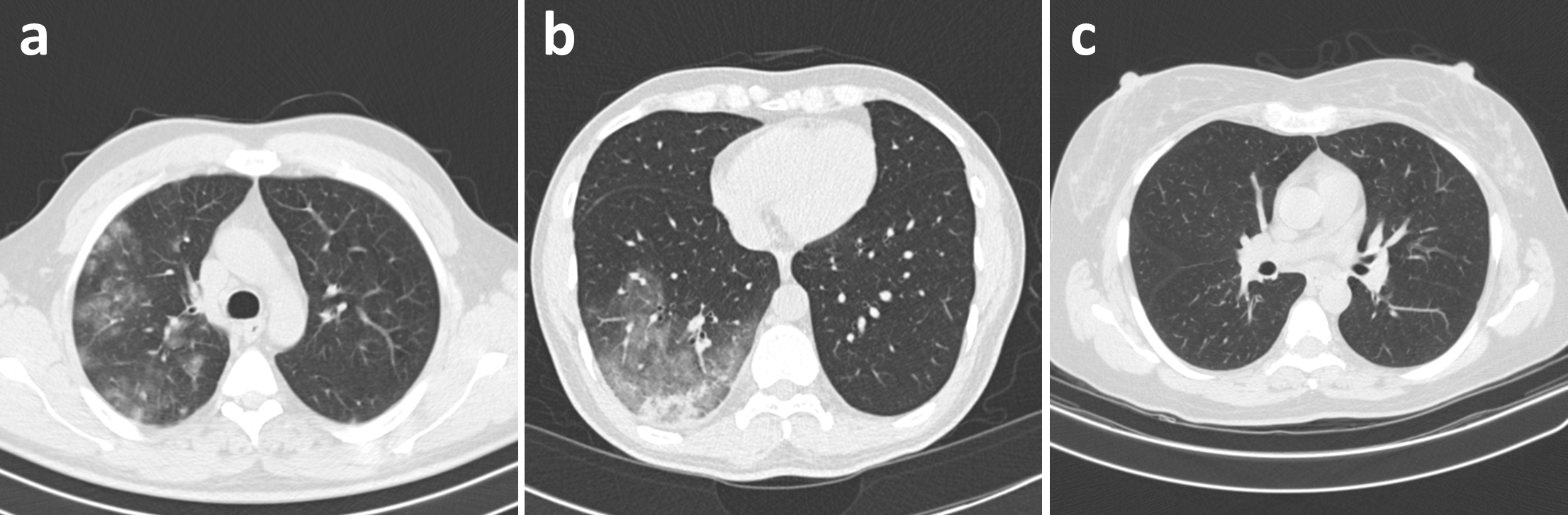}
        \caption{Example chest CT images from the COVIDx~CT-3 dataset. (a) a COVID-19 case, (b) a CAP case, (c) a normal control.}
        \label{fig:examples}
        \vspace{-0.1in}
    \end{figure}

\label{sec:sources}
    In this study, we carefully processed and curated CT images from several data sources from around the world which were collected using a variety of CT scanners and protocols. The resulting patient cohort consists of patient cases collected by the following organizations and initiatives: (1) China National Center for Bioinformation (CNCB)~\cite{cncb}, (2) National Institutes of Health Intramural Targeted Anti-COVID-19 (ITAC) Program~\cite{TCIACOVID, TCIA}, (3) COVID-CTset~\cite{COVIDCTset}, (4) Integrative CT Images and Clinical Features for COVID-19 (iCTCF)~\cite{HUST}, (5) COVID-19 CT Lung and Infection Segmentation initiative (COVID-19-CT-Seg)~\cite{COVID-19-SegBenchmark}, (6) Lung Image Database Consortium and Image Database Resource Initiative (LIDC-IDRI)~\cite{TCIA,LIDC,LIDC2}, (7) Radiopaedia collection~\cite{radiopaedia}, (8) MosMedData~\cite{MosMedData}, (9) Stony Brook University~\cite{TCIA,stonybrook}, (10) Study of Thoracic CT in COVID-19 (STOIC)~\cite{STOIC}, and (11) COVID-CT-MD~\cite{COVIDCTMD}.
    
    Each patient is associated with one of three possible infection types: (a) COVID-19, (b) community-acquired pneumonia (CAP), or (c) normal controls, with Figure~\ref{fig:examples} illustrating an example of each infection type. For CT volumes which were not labelled at the image level, labels were obtained in one of three ways: (1) segmentation-based labelling, where ground-truth infection masks were used to identify abnormal CT slices, (2) non-expert manual labelling, where non-experts manually labelled CT slices with obvious abnormalities, or (3) model-based automatic labelling, where a pre-trained model~\cite{Gunraj2022} was used to identify CT slices with high CAP or COVID-19 confidence. For all three methods, the selected CT slices were assigned the same labels as their respective patients.

\section{Results}   
    \subsection{Demographics}
    
    Table~\ref{tab:demographics} shows the demographics of the COVIDx CT-3 dataset. Examining the countries of origin, we see that Chinese patients dominate the data, accounting for 42.2\% of the patient cohort. Additionally, we see that the vast majority of patients (85.9\%) were imaged in one of four countries (China, France, Russia, or Iran), illustrating a severe geographical bias towards Asian and European patients. Additionally, we see that age and sex are unknown for more than half of the patient cohort. Of the patients with known ages, the majority (94.3\%, or 45.9\% of all patients) fall within the age range 30-89, indicating that young patients ($\leq$ 29) and very old patients ($\geq$ 90) may be underrepresented. Of the patients with known sexes, there is a reasonably even split between male and female patients (27.0\% and 22.1\%, respectively).

    \begin{table}[!ht] 
        \centering
        \caption{Summary of demographics for the patient cohort examined in this study.}
        \label{tab:demographics}
        \begin{tabular}{|l|l|l|l||l|l|}
            \hline
            \multicolumn{4}{|l||}{\textbf{Country}} & \multicolumn{2}{l|}{\textbf{Age}}\\
            \hline
            Unknown & 702 (11.6\%) & Scotland & 1 (0.02\%) & Unknown & 3120 (51.4\%) \\
            China & 2563 (42.2\%) & Peru & 1 (0.02\%) & 0-9 & 17 (0.3\%) \\
            France & 1176 (19.4\%) & Lebanon & 1 (0.02\%) 
 & 10-19 & 24 (0.4\%) \\
            Russia & 756 (12.5\%) & England & 1 (0.02\%)  & 20-29 & 114 (1.9\%) \\
            Iran & 718 (11.8\%) & Turkey & 1 (0.02\%) & 30-39 & 428 (7.1\%) \\
            USA & 129 (2.1\%) & Belgium & 1 (0.02\%) & 40-49 & 467 (7.7\%) \\
            Australia & 7 (0.12\%) & Azerbaijan & 1 (0.02\%) & 50-59 & 576 (9.5\%) \\
            Algeria & 5 (0.08\%) & Afghanistan & 1 (0.02\%) & 60-69 & 606 (10.0\%) \\
            Italy & 3 (0.05\%) & Ukraine & 1 (0.02\%) & 70-79 & 403 (6.6\%) \\
            \cline{1-4}
            \multicolumn{4}{|l||}{\textbf{Sex}} & 80-89 & 302 (5.0\%) \\
            \cline{1-4}
            \multicolumn{2}{|l|}{Unknown} & \multicolumn{2}{l||}{3091 (50.9\%)} & 90-99 & 11 (0.2\%) \\
            \multicolumn{2}{|l|}{Male} & \multicolumn{2}{l||}{1639 (27.0\%)} & & \\
            \multicolumn{2}{|l|}{Female} & \multicolumn{2}{l||}{1338 (22.1\%)} & & \\
            \hline
        \end{tabular}
    \end{table}

    \subsection{Data Splits}
    
    Table~\ref{tab:splits} shows the distribution of CT slices and patients amongst the training, validation and test sets. The data is split into training, validation, and test sets with fractions of approximately 84\%, 8\%, and 8\%, respectively. This is due to the fact that data labelled automatically or by non-experts is included solely in the training set, while the validation and test sets were both labelled manually by experts. Additionally, we observe a class imbalance where COVID-19 images represent 73.4\% of the data and normal and pneumonia images represent 16.6\% and 10.0\% of the data, respectively. Notably, this imbalance is primarily seen in the training set, while the validation and test sets both have ratios of approximately 2.2:1:1 between Normal, CAP, and COVID-19 images, respectively.

    \begin{table}[!ht]
        \caption{Distribution of chest CT slices and patient cases (in parentheses) by data split and infection type in COVIDx~CT-3.}
        \label{tab:splits}
        \centering
        \begin{tabular}{lcccc}
            \toprule
            & \multicolumn{3}{c}{\textbf{Infection Type}} & \\
            \cmidrule(lr){2-4}
            \textbf{Data Split} & Normal & CAP & COVID-19 & \textbf{Total} \\
            \midrule
            Training & 35,996 (321) & 26,970 (592) & 300,733 (4,092) & 363,699 (5,005)\\
            \midrule
            Validation & 17,570 (164) & 8,008 (202) & 8,147 (194) & 33,725 (560)\\
            \midrule
            Test & 17,922 (164) & 7,965 (138) & 7,894 (201) & 33,781 (503)\\
            \midrule
            \textbf{Total} & 71,488 (649) & 42,943 (932) & 316,774 (4,487) & 431,205 (6,068)\\
            \bottomrule
        \end{tabular}
        \vspace{-0.1in}
    \end{table}

    \subsection{Benchmarks}
    Table~\ref{tab:benchmarks} provides performance measures for a variety of network architectures on the COVIDx~CT-3 test dataset, as previously presented in~\cite{Gunraj2022}. These results provide reference performance targets for new networks trained on the COVIDx~CT-3 dataset.

    \begin{table}[!ht]
        \caption{Comparison of parameters, floating-point operations, accuracy (image-level), COVID-19 sensitivity (image-level), and COVID-19 positive predictive value (image-level) for benchmark networks on the COVIDx~CT-3 test dataset. Best results highlighted in \textbf{bold}.}
        \medskip
        \label{tab:benchmarks}
        \centering
        \begin{tabular}{lccccc}
            \toprule
            \textbf{Network} & \textbf{Param. (M)} & \textbf{FLOPs (G)} & \textbf{Acc. (\%)} & \textbf{Sens. (\%)} & \textbf{PPV (\%)}\\
            \midrule
            SqueezeNet~\cite{squeezenet} & 0.74 & 8.09 & 98.7 & 97.7 & \textbf{98.1}\\
            MobileNetV2~\cite{mobilenetv2} & 2.23 & 3.33 & \textbf{99.0} & 98.5 & 98.0\\
            EfficientNet-B0~\cite{efficientnet} & 4.05 & 4.07 & \textbf{99.0} & \textbf{99.1} & 98.0\\
            NASNet-A-Mobile~\cite{nasnet} & 4.29 & 5.94 & 98.8 & 98.7 & 96.9\\
            COVID-Net~CT~L~\cite{Gunraj2020} & 1.40 & 4.18 & 98.4 & 98.1 & 96.1\\
            COVID-Net~CT~S~\cite{Gunraj2022} & \textbf{0.45} & \textbf{1.94} & 98.3 & 97.3 & 96.3\\
            \bottomrule
        \end{tabular}
        \vspace{-0.1in}
    \end{table}
    
\section{Discussion}
    
    Given the demographic imbalances of the COVIDx~CT-3 dataset, algorithms and strategies accounting for such imbalances (e.g., balanced loss functions, data sampling and re-balancing methods, etc.) may be an interesting area of exploration. Additionally, due to the large number of patients with unknown ages or sexes, it is difficult to ascertain the true age and sex distributions.
    
    Systems trained on the COVIDx~CT-3 dataset may require careful re-balancing of the classes in order to address the class imbalance in the training set. Additionally, evaluation of such systems should take into account the imbalances in the validation and test sets by examining balanced metrics, such as per-class precision and recall.

\section*{Potential Negative Societal Impact}
    While the motivation behind the release of this large-scale benchmark dataset is to support researchers, clinicians, and citizen data scientists in advancing this field, a potential negative societal impact from this release of is the misuse of the collected data. More specifically, misuse of collected data can occur if the benchmark dataset is used to build machine learning algorithms for the purpose of forecasting future medical expenses for individual patients, which insurance companies may use to adjust insurance premiums.

    Furthermore, use of this dataset to build predictive models for clinical use can have a negative impact if such models are not properly validated. In particular, the use of automated and non-expert labelling methods in the construction of this dataset inevitably introduces a degree of label noise which may affect the performance and behaviour of any developed models. To avoid this, models built using this dataset should be validated on real-world clinical data, should not be used in clinical settings without expert oversight, and should not be used to make final diagnostic decisions.

\section*{Acknowledgements}
    We would like to thank the Canada Research Chairs program and the the Natural Sciences and Engineering Research Council of Canada (NSERC).

\bibliographystyle{unsrtnat}
\bibliography{references}

\end{document}